\documentclass{article}

\usepackage{arxiv}

\usepackage[utf8]{inputenc} 
\usepackage[T1]{fontenc}    
\usepackage{hyperref}       
\hypersetup{
	colorlinks   = true, 
	urlcolor     = blue, 
	linkcolor    = blue, 
	citecolor   = blue 
}
\usepackage{url}            
\usepackage{booktabs}       
\usepackage{amsfonts}       
\usepackage{nicefrac}       
\usepackage{microtype}      
\usepackage{lipsum}
\usepackage{graphicx}
\usepackage{amsmath}
\usepackage{amsfonts}
\usepackage{amssymb}
\usepackage{multirow}
\usepackage{float}
\usepackage{adjustbox}
\usepackage{booktabs}
\usepackage{lscape}
\usepackage{enumitem, array}
\usepackage{xcolor}
\usepackage[round,authoryear]{natbib}
\usepackage{algorithm2e}
\usepackage{subcaption}
\SetAlgorithmName{Procedure}{procedure}{List of Procedures}

\usepackage{tikz}
\usepackage{bm}
\usepackage{relsize}

\usetikzlibrary{positioning}

\title{Towards a Statistical Validation of the Critical Wave Groups Method for Free-Running Vessels in Beam Seas}

\author{
	Kevin M. Silva$^{1,\star}$ and Kevin J. Maki$^{2}$\\
	\\
	$^1$Naval Surface Warfare Center Carderock Division, USA\\
	$^2$Department of Naval Architecture and Marine Engineering, The University of Michigan, USA\\
	$^\star$Corresponding author: \texttt{kevin.m.silva14.civ@us.navy.mil} \\
}

\begin{document}
	\maketitle
	
\begin{abstract}
Research on the statistics of extreme events using deterministic wave group methods has largely been simplified to vessels at zero or constant speed and heading. In contrast, free-running vessels move with six degrees-of-freedom (6-DoF), leading to more complex and varied extreme response events. This paper details the extension of the Critical Wave Groups (CWG) method to free-running vessels and demonstrates that the method produces probability calculations comparable to those from a limited Monte Carlo dataset for a vessel in beam seas. This research is a critical first step in the formal validation of this free-running implementation of the CWG method.
\end{abstract}

\section*{Introduction}
Throughout their operational lifetime, naval vessels experience a wide variety of waves that can induce extreme responses. Proper evaluation of these extreme events is critical not only for designing new vessels but also for operating the current fleet. These extreme response events include capsizing, broaching, and large loads that can lead to catastrophic failures and possible loss of vessels, equipment, and personnel. It is critical to not only quantify these types of events probabilistically but also observe the extreme ship responses quantitatively so that the underlying mechanisms causing them can be studied and understood.

Due to the stochastic nature of the ocean, direct calculation of extreme event statistics with Monte Carlo approaches are impractical because of the long exposure times required. To combat the need for long exposure times, several probabilistic methodologies have been implemented to quantify extreme events. Some examples include the Peaks Over Threshold (POT) \citep{Campbell2010a} and the Envelope Peaks Over Threshold (EPOT) \citep{Campbell2010b, Belenky2011b, Campbell2023} methods. In addition to the extrapolation methods is the split-time method which utilizes a distance to danger methodology to identify extreme events \citep{Belenky1993, Belenky2010, Weems2020, Belenky2024}. 

However, these extreme event probabilistic methods do not provide physically realizable events and are more concerned with the probabilistic calculation. Wave group and wave episode methods are more focused on the observation of extreme events and the probabilistic quantification is a byproduct of observing a sufficient amount of tailored events.

Some of the wave group and wave episode methods include the Design Loads Generator (DLG) \citep{Alford2008, Alford2011, Kim2012}, Gaussian Process Regression (GPR) sequential sampling with reduced-order wave groups \citep{Mohamad2018, Sapsis2021}, the use of GPR and Karhunen-Loève (KL) expansion of the wave field and ship motion \citep{Guth2022}, and the Critical Wave Groups (CWG) method \citep{Themelis2007, Anastopoulos_etal2016, Anastopoulos2016, Anastopoulos2017, Anastopoulos2019}.

The majority of research in observing and quantifying the statistics of extreme events with deterministic wave group methods has involved simplifications such as only considering zero speed or constant speed and heading. However, free-running vessels travel with six degrees-of-freedom (6-DoF) and surge, sway, and yaw motions in conjunction with propeller and rudder forces can lead to extreme ship response events and even capsizing. Extreme events are not only triggered by large wave events but also depend on how the ship encounters the waves and the ship’s position in the wave field. A primary challenge in applying deterministic wave group methods to free-running vessels is ensuring that the ship’s encounter with the wave group is repeatable and consistent across different wave groups and encounter conditions in order to perform probabilistic assessments accurately.

Of the popular wave group and wave episode methodologies, only the CWG method has been extended to a free-running implementation in \cite{Silva2022snh} and \cite{Silva2023} for simulations utilizing computational fluid dynamics (CFD). Although previous research was able to demonstrate the methodology and feasibility for evaluating extreme events with CFD, the CFD simulations proved too computationally expensive for a statistical validation of the framework. 

The objective of the present paper is to build upon previous research and move towards a statistical validation of the free-running implementation of the CWG method developed in \cite{Silva2022snh} and \cite{Silva2023} through potential flow simulations utilizing the Large Amplitude Motion Program (LAMP) \citep{Lin1994, Lin2007}. The use of a computational tool that is less expensive than CFD will allow for a larger dataset to compare the probability calculations for the CWG method. This paper will describe the CWG method as used in this research, the extension to free-running vessels, and detail a case study that lays the foundation for a complete statistical validation of the CWG method for free-running vessels.

\section*{Critical Wave Groups Method}

The CWG method is a probabilistic framework for observing extreme ship response events and quantifying the probability of their occurrence utilizing targeted deterministic wave groups. The CWG method was initially developed for regular waves in \cite{Themelis2007} and later extended to irregular waves in \cite{Anastopoulos_etal2016, Anastopoulos2016, Anastopoulos2017, Anastopoulos2019}. This paper relies heavily on the methodology formulated in \cite{Anastopoulos2019} and is further described in detail in \cite{Silva2023}.

\subsection*{Probability of Exceedance}

The CWG method is focused on calculating the probability that a given response $\phi$ exceeds a specified threshold of interest $\phi_{\rm crit}$. The CWG method assumes that the probability that $\phi$ exceeds $\phi_{\rm crit}$ is a function of the wave group that acts on the ship, and the motion state of the vessel at the moment the wave group is encountered, referred herein as the \emph{encounter condition}, but also referred to in the CWG literature as the \emph{initial condition}. For sample spaces of arbitrary parameterized wave groups, $G$, and encounter conditions, $E$, a probability of exceedance can be generally described as:

\begin{equation}
	p \left[ \phi > \phi_{\rm crit} \right] \ = \ 
	\int \textbf{1}_{\Theta\left(g,e\right) > \phi_{\rm crit}}
	f_{G,E}\left(g,e\right) dg\,de
	\label{eq:probexceed_gen}
\end{equation}

where $f_{G,E}\left(g,e\right)$ is the joint probability density function (PDF) of wave groups and encounter conditions, $\Theta$ is a mapping describing the absolute maximum response of the ship for a given wave group and encounter condition ($\theta = \Theta\left(g,e\right)$), and $\textbf{1}_{\Theta\left(g,e\right) > \phi_{\rm crit}}$ is an indicator function that denotes a result of one when a particular wave group/encounter condition pair exceeds a given threshold, and an output of zero when it does not. To evaluate Eqn.~\eqref{eq:probexceed_gen}, predictions of the ship response due to different wave groups and encounter conditions are needed to develop the mapping $\Theta$. Due to the stochastic nature of the ocean, there are infinite possible combinations of wave groups and encounter conditions which makes the complete evaluation of $\Theta$ expensive.

The CWG method focuses on defining the boundary created by the indicator function in Eqn.~\eqref{eq:probexceed_gen}. The corresponding wave groups and encounter conditions along this indicator function boundary lead to responses that cause a near-exceedance of the specified response threshold $\phi_{\rm crit}$. These wave groups are referred to as the \emph{critical wave groups}.

Eqn.~\eqref{eq:probexceed_gen} does not make any assumption about the shape of the wave groups or the nature of the encounter conditions. To practically solve Eqn.~\eqref{eq:probexceed_gen}, the encounter conditions must be discretized and the wave groups are parameterized by their shape to create a set of mutually exclusive and collectively exhaustive wave groups. With an arbitrary shaping parameter $q$ and encounter conditions $ec_k$, Eqn.~\eqref{eq:probexceed_gen} can be rewritten as:

\begin{equation}
p \left[ \phi > \phi_{\rm crit} \right] =
\sum\limits_{k}{} 
p \left[ \bigcup_q wg_{k,q}, ec_k \right]
\label{eq:probexceed_param1}
\end{equation}

where $ec_k$ is the $k^{th}$ encounter condition, and $wg_{k,q}$ are all the wave groups for the $k^{th}$ encounter condition and the $q^{th}$ wave group shape that lead to the response exceeding a specified threshold.

In this paper, wave groups are classified in terms of their run length $j$ and wave period groupings in $m$. The parametrization utilizing $j$ and $m$ results in wave groups that contain the same number of waves and each wave in the group is within a similar range of period. By introducing the quantities $j$ and $m$, Eqn.~\eqref{eq:probexceed_param1} can be further expanded as:

\begin{equation}
p \left[ \phi > \phi_{\rm crit} \right] =
\sum\limits_{k}{} 
\sum\limits_{m}{} 
p \left[ \bigcup_j wg_{m,j}^{(k)} \right] \times
p \left[ ec_k \right]
\label{eq:probexceed_param4}
\end{equation}

where $wg_{m,j}^{(k)}$ are all the wave groups leading to a threshold exceedance with $j$ waves, wave periods in the $m^{th}$ wave period range, and the $k^{th}$ encounter condition. By assuming that for sufficiently large responses, the wave groups $wg_{m,j}^{(k)}$ are rare events and statistically independent of one another, Eqn.~\eqref{eq:probexceed_param4} can be further rewritten as: 

\begin{equation}
p \left[ \phi > \phi_{\rm crit} \right] =
\sum\limits_{k}{} 
\sum\limits_{m}{}
\left(	
1-
\prod_{j}\left( 1-p \left[ wg_{m,j}^{(k)} \right]\right)
\right)
\times
p \left[ ec_k \right]
\label{eq:probexceed_param}
\end{equation}

where the probability of exceedance is now broken into two separate components: the probability of the$k^{th}$ encounter condition $p \left[ ec_k \right]$ and the probability of wave group exceedance $p \left[ wg_{m,j}^{(k)} \right]$, for the $m^{th}$ wave period range, $j$ waves in the group, and the$k^{th}$ encounter condition. Further details on the derivation of the probability of exceedance can be found in \cite{Anastopoulos2019} and \cite{Silva2023}.

\begin{figure}[H]
    \centering
 \begin{subfigure}{0.49\textwidth}
     \centering
    \includegraphics[width=0.99\textwidth]{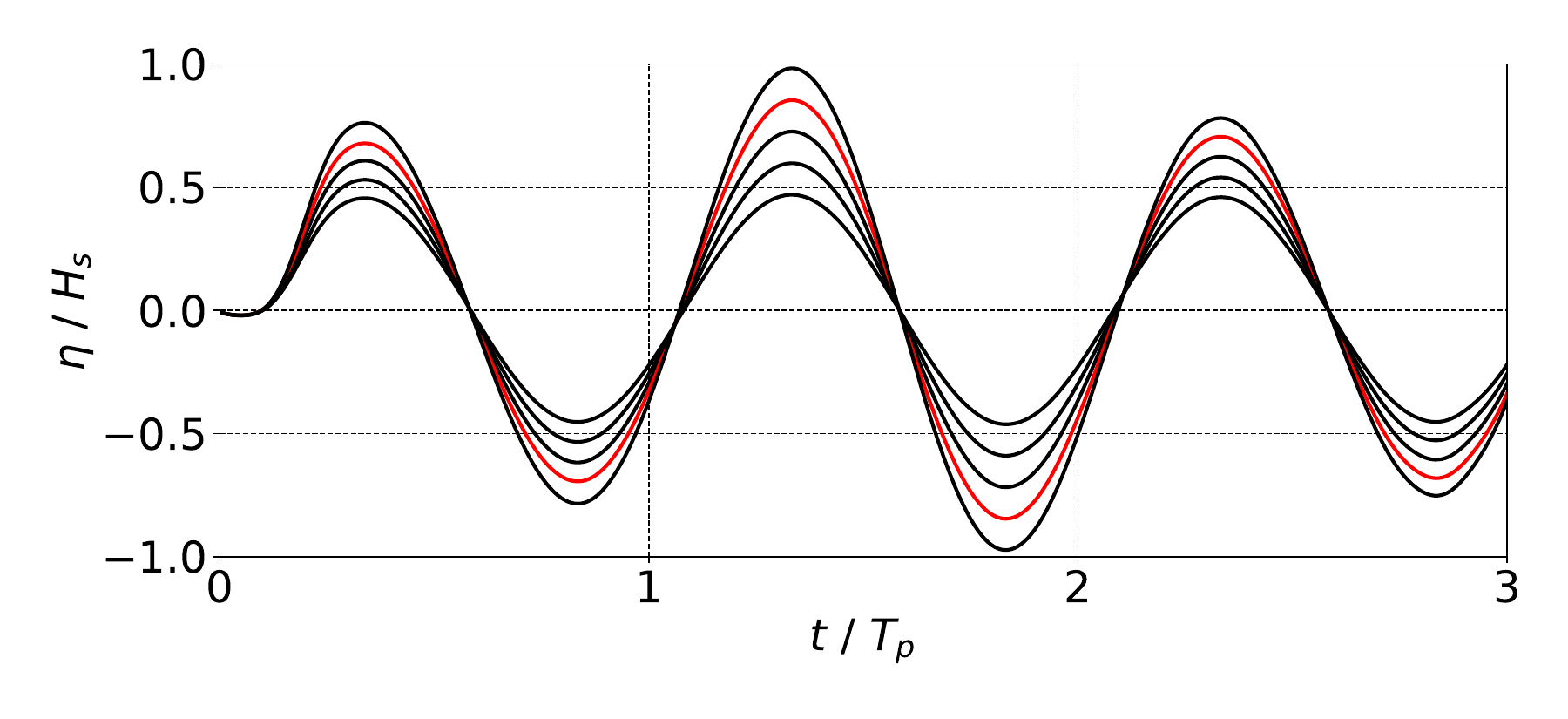}
     \caption{Wave Groups}
 \end{subfigure}
 \begin{subfigure}{0.49\textwidth}
     \centering
    \includegraphics[width=0.99\textwidth]{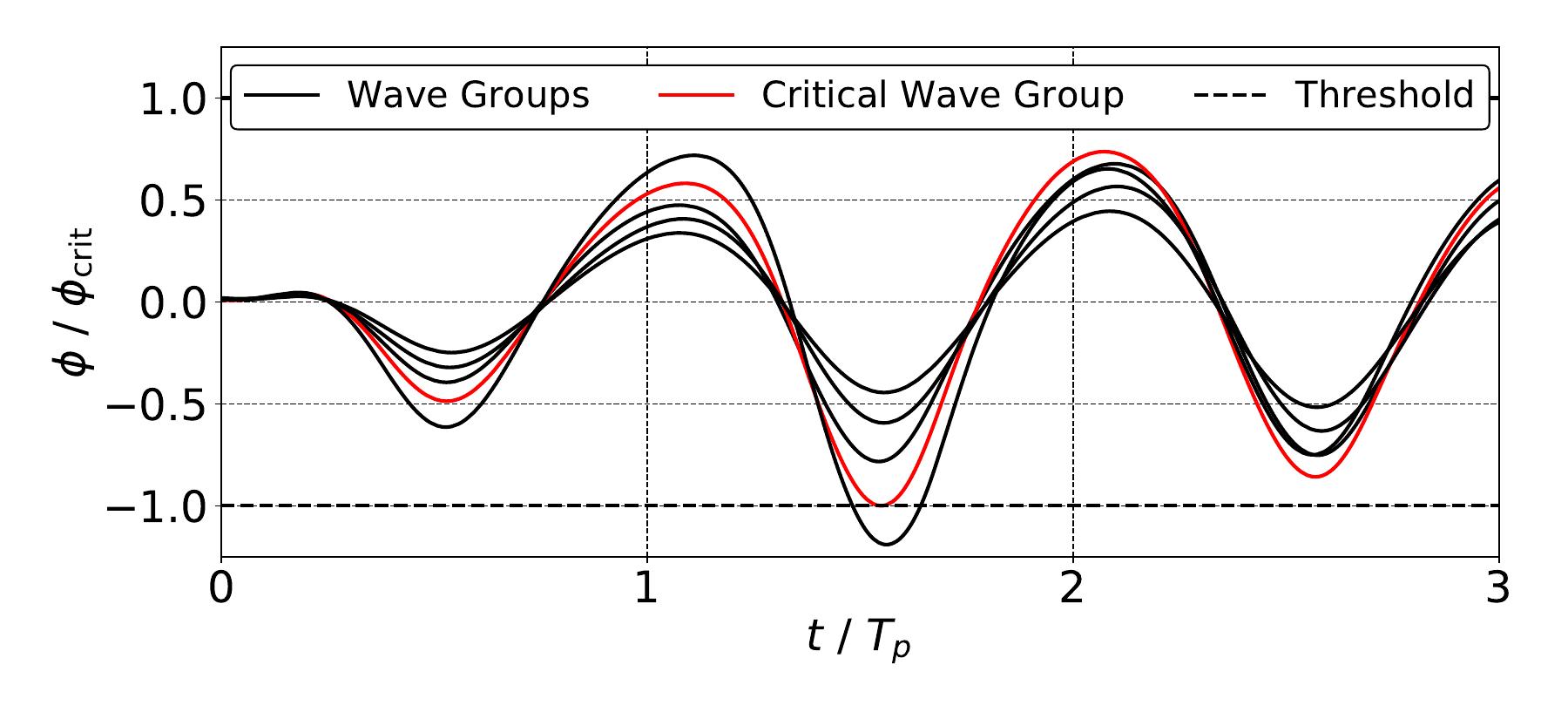}
     \caption{Response}
 \end{subfigure}
    \caption{Identification of a critical wave group for a given set of wave groups with similar shapes.}
\label{fig:crit}
\end{figure}

Fig.~\ref{fig:crit} shows an example of a series of wave elevation $\eta$ wave groups with the same $j$ and $m$, but with varying wave heights. $H_s$ corresponds to the significant wave height, $t$ is time, and $T_p$ is the peak wave period. The resulting ship responses are also shown in Fig.~\ref{fig:crit}. By incrementally increasing the wave group height, eventually a wave group will result in a ship response that meets the specified threshold as shown in red in Fig.~\ref{fig:crit}. Any wave group that is larger in height than the critical wave group, is assumed to also exceed the response threshold. Therefore, the probability of exceedance calculation requires knowledge of the critical wave group and does not require knowledge of the smaller wave groups.

\subsection*{Wave Group Construction}

Wave groups in this work are constructed in the same manner formulated in \cite{Anastopoulos2019} utilizing Markov chains that consider the statistical relationship between the heights and periods of successive waves. Given the height and period of the largest wave in a wave group, the Markov chain methodology can be employed to predict the most expected preceding and following wave. The description of each wave in the sequence only depends on the closest successive wave because of the Markov chain’s memoryless property. Therefore, a wave group with $j$ waves can be fully described by only the height $H_c$ and period $T_c$ of the largest wave in the group.

Eqn.~\eqref{eq:tpred} and \eqref{eq:ptn} demonstrate how the period of the most expected successive wave $\overline{t_n}$ can be calculated based off the current wave height $h_{n-1}$ and period $t_{n-1}$, as well as the as well as the transition kernel $f_{T_n |H_{n-1},T_{n-1}} \left( t_n | h_{n-1}, t_{n-1}  \right)$. The transition kernel is a conditional probability density function (PDF) that represents a relationship between the succeeding wave period $t_n$ and the current wave's height and period. In this paper, transition kernels are found through generating large datasets of random irregular waves and developing the statistical relations with the actual data as opposed to spectral methods explained in \cite{Anastopoulos_etal2016}.

\begin{equation}
\overline{t_n} \ = \  \frac{1}{p_{T_n}} \int\limits_{T_{cr,m}} t_n 
f_{T_n |H_{n-1},T_{n-1}} \left( t_n | h_{n-1}, t_{n-1}  \right) dt_n
\label{eq:tpred}
\end{equation}

\begin{equation}
p_{T_n} \ = \ 
\int\limits_{T_{cr,m}}
f_{T_n |H_{n-1},T_{n-1}} \left( t_n | h_{n-1}, t_{n-1}  \right) dt_n
\label{eq:ptn}
\end{equation}

Eqn.~\eqref{eq:hpred} describes the prediction of the most expected wave height wave $\overline{h_n}$ in a sequence based on the most expected wave period predicted with Markov chains in Eqn.~\eqref{eq:tpred} and \eqref{eq:ptn} and the current wave’s height $h_{n-1}$ and period $t_{n-1}$. 

\begin{equation}
\overline{h_n} \ = \ \int\limits_{0}^{\infty} h_n 
f_{H_n |T_n,H_{n-1},T_{n-1}} \left( h_n | \overline{t_n}, h_{n-1}, t_{n-1}  \right) dh_n
\label{eq:hpred}
\end{equation}

The Markov chain procedure produces deterministic predictions of the heights and periods of the individual waves within a group based on the height $H_c$ and period $T_c$ of the largest wave in the group and the number of waves in the group $j$. 

\begin{figure}[H]
	\centering
	\includegraphics[width=0.7\textwidth]{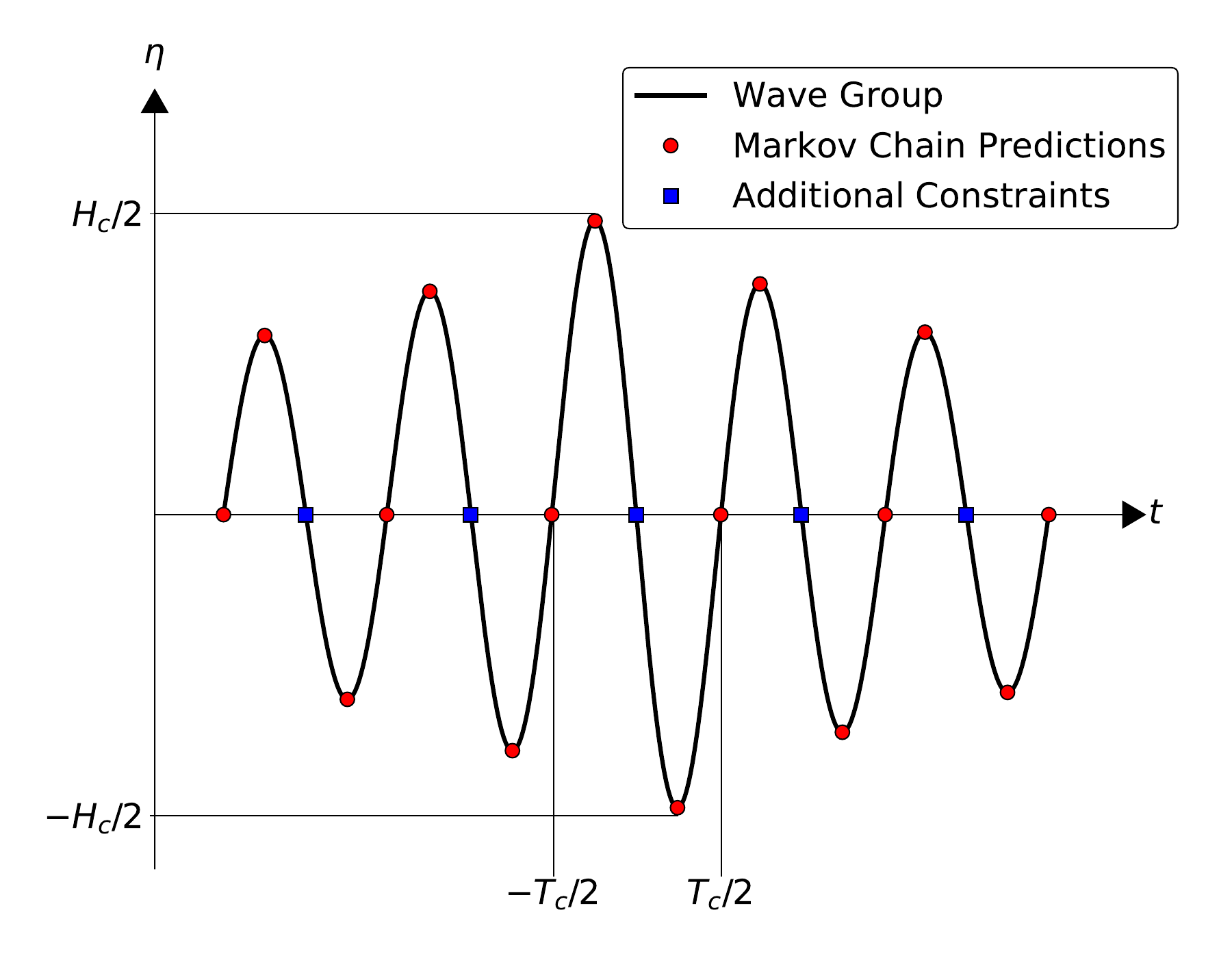}
	\caption{Markov chain construction of wave groups and additional geometric constraints (figure adapted from \cite{Anastopoulos2016}).}
	\label{fig:wgconstraint}
\end{figure}

However, continuous temporal representations of the wave groups are required for evaluation within hydrodynamic simulation tools or experimental wave basins. Fig.~\ref{fig:wgconstraint} shows the generation of a continuous temporal representation of a deterministic wave group elevation $\eta$ for time $t$. The red circles correspond to the Markov chain predictions of wave height and periods for each wave in the group utilizing Eqn.~\eqref{eq:tpred} through \eqref{eq:hpred}. Additional constraints are denoted by blue squares through the assumption that each wave in the group is symmetry in terms of its height and period. For each wave in the group, the crest and trough are assumed to each be half the wave height, and the time derivative of wave elevation is set to zero at the crest and trough to ensure they are peaks. Also, the crest and trough occur at the center of the time interval defined by the successive zero-crossings, and the zero-crossings occur at instances of half of the current wave period. With the Markov chain predictions and geometric constraints, a continuous time representation is created through trigonometric interpolation to define the wave group as a series of Fourier components.

\begin{figure}[H]
	\centering
	\includegraphics[width=0.7\textwidth]{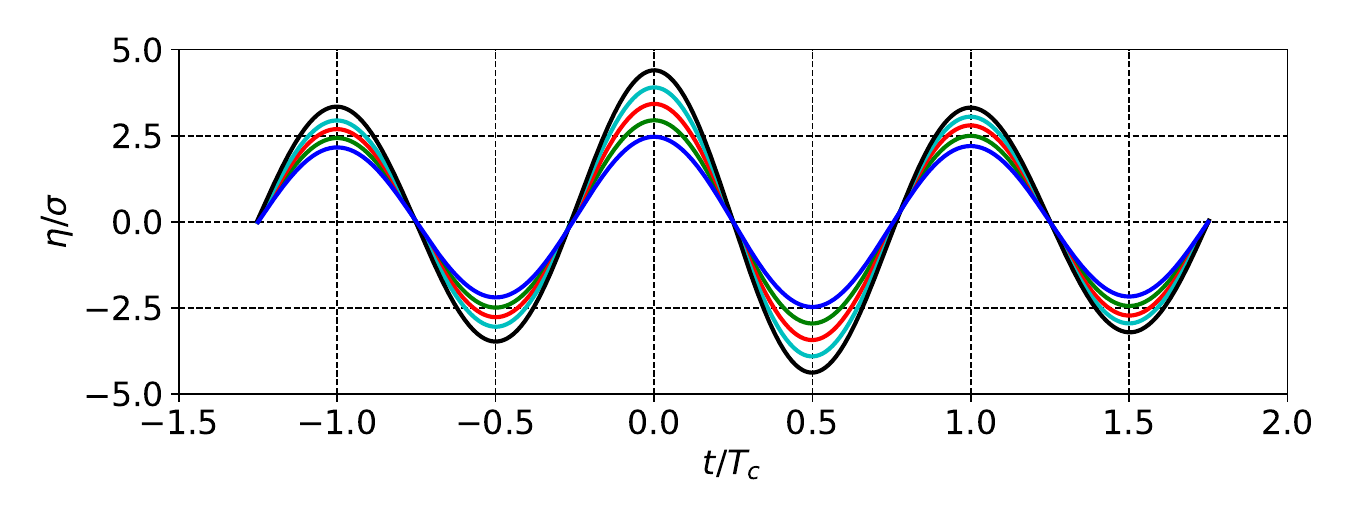}
	\caption{Ensemble of wave groups with the same $T_c$, $j$, and $H_c$ = 5$\sigma$, 6$\sigma$, 7$\sigma$, 8$\sigma$, and 9$\sigma$.}
	\label{fig:cwg_ensembleWG}
\end{figure}

Fig.~\ref{fig:cwg_ensembleWG} shows an example of wave groups with the same $T_c$ and $j$ with varying values of $H_c$ to showcase the influence the largest wave in the group has on the succesive waves. $\sigma$ is the standard deviation of wave height from the random wave observations.

\subsection*{Wave Group Probability}

Eqn.~\eqref{eq:probexceed_gen} through \eqref{eq:hpred} show the overall probability of exceedance calculation, the identification of critical wave groups, and the construction of the deterministic wave groups utilizing Markov chains. However, solution of Eqn.~\eqref{eq:probexceed_param} requires calculation of the probability that a wave group exceeds the critical wave group $p \left[ wg_{m,1}^{(k)} \right]$ for the for the $m^{th}$ wave period range, $j$ waves in the group, and the $k^{th}$ encounter condition. Eqn.~\eqref{eq:probwg} shows the probability of wave group exceedance, where the probability that a wave group is larger than the critical wave group is the joint probability that the waves in a wave group $H_j$ are larger than the corresponding waves in the critical wave group $\mathbf{h}_{cr,j,m}^{(k)}$ and that the period of the waves in the wave group $T_j$ are within the wave period range $T_{cr,m}$.

\begin{equation}
p \left[ wg_{m,j}^{(k)} \right] \ = \ 
p \left[ \mathbf{H}_j > \mathbf{h}_{cr,j,m}^{(k)},\mathbf{T}_j \in T_{cr,m} \right]
\label{eq:probwg}
\end{equation}

Following the research of \cite{Anastopoulos2019}, $p \left[ wg_{m,j}^{(k)} \right]$ can be calculated utilizing Equations \eqref{eq:probwg_final} through \eqref{eq:p0} utilizing the Markov chain relationship. A more detailed derivation and explanation can be found in \cite{Silva2023}.

\begin{equation}
	p \left[ wg_{m,j}^{(k)} \right] \ = \ 
	p_0 \ \times \ \prod \limits_{n=2}^{j} \frac{p_1^{(n)} \ \times \ p_2^{(n)}}
	{p_{01}^{(n-1)} \ \times \ p_{02}^{(n-1)}}
	\label{eq:probwg_final}
\end{equation}

\begin{equation}
	p_1^{(n)} \ = \
	\int\limits_{T_{cr,m}} \int\limits_{T_{cr,m}} 
	f_{T_n, T_{n-1} |\mathbf{H}_{n}} \left( t_n, t_{n-1} | \mathbf{H}_{n} > \mathbf{h}_{cr,n}  \right) dt_n dt_{n-1} 
	\label{eq:p1}
\end{equation}

\begin{equation}
	p_2^{(n)} \ = \
	\int\limits_{h_{cr,n}}^{+\infty} \int\limits_{h_{cr,n}}^{+\infty} 
	f_{H_n, H_{n-1}} \left( h_n, h_{n-1} \right) dh_n dh_{n-1} 
	\label{eq:p2}
\end{equation}

\begin{equation}
	p_{01}^{(n)} \ = \
	\int\limits_{T_{cr,m}}
	f_{T_n |\mathbf{H}_{n}} \left( t_n | \mathbf{H}_{n} > \mathbf{h}_{cr,n}  \right) dt_n
	\label{eq:p01}
\end{equation}

\begin{equation}
	p_{02}^{(n)} \ = \
	\int\limits_{h_{cr,n}}^{+\infty}
	f_{H_n} \left( h_n \right) dh_n
	\label{eq:p02}
\end{equation}

\begin{equation}
	p_{0} \ = \
	p_{01}^{(1)} \ \times \ p_{02}^{(1)}
	\label{eq:p0}
\end{equation}

\section*{Natural Initial Condition}

Previous implementations of the CWG method in \cite{Themelis2007, Anastopoulos_etal2016, Anastopoulos2016, Anastopoulos2017, Anastopoulos2019} all considered one degree-of-freedom ordinary differential equation (ODE) models of roll where the entire dynamic state when encountering the wave group can be easily defined through an initial condition. However, in reality and higher fidelity hydrodynamic simulation tools, the entire dynamical state of the vessel and the surrounding fluid cannot be fully prescribed instantaneously with an initial condition. Therefore, the natural initial condition was developed in \cite{Silva2021oe, Silva2024, Silva2021stab} for implementing the the CWG method with CFD for the extreme roll of a two-dimensional midship section and then later extended to free-running vessels in \cite{Silva2022snh} and \cite{Silva2023}. The main idea behind the natural initial condition is for the encounter conditions to be achieved naturally as opposed to forcefully prescribing the state of the vessel as an initial condition. The natural initial condition achieves the encounter condition of interest by embedding deterministic wave groups into an ensemble of previously observed irregular seaways that will naturally produce different encounter conditions as a ship reaches the wave group of interest. This methodology avoids the issues associated with explicitly prescribing initial conditions, and instead allows for the fluid flow and ship responses to develop naturally. Additionally, prescribing the encounter conditions in this manner preserves the integrity of the CWG methodology developed for an ODE, while making the method accessible for higher-fidelity numerical hydrodynamic simulation tools and physical experiments as well. 

\begin{figure}[H]
	\centering
	\includegraphics[width=0.7\textwidth]{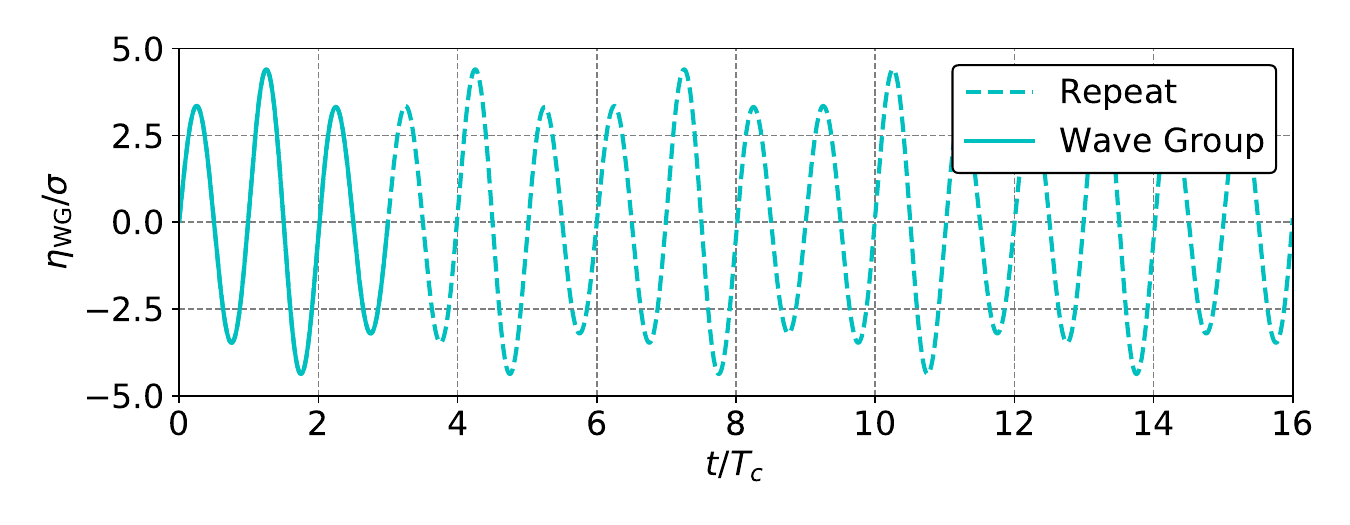}
	\caption{Representation of deterministic wave group at origin with Fourier components.}
	\label{fig:ic_etaWG}
\end{figure}

Consider a single realization of a free-running vessel starting traveling through a random seaway that evolves in space and time. Throughout the realization, a specific encounter condition of interest occurs at time $t_e$. The waves experienced by the vessel in the encounter frame can be approximated as the wave elevation $\eta_{\rm IP}^{\rm (E)} (t)$, which up until time $t_e$ is referred to as the irregular prelude. The ship speed $$U_e = x_e/t_e$$ can also be estimated based on the ship traveling a distance $x_e$ in time $t_e$. The deterministic wave groups are defined in the earth-fixed frame through Fourier components which results in a wave group elevation time-history at the origin ($x,y = 0,0$):

\begin{equation}
\eta_{\rm WG} \left(  t \right) \  = \ \sum_{f}^{} a_{g_f} cos \left( \omega_{g_f} t \ + \ \phi_{g_f} \right)
\label{eq:ic_wg}
\end{equation}

where $a_{g_f}$, $\omega_{g_f}$, and $\phi_{g_f}$ are the Fourier amplitudes, frequencies, and phases describing the deterministic wave group. Fig.~\ref{fig:ic_etaWG} shows the wave elevation at the origin described by the Fourier components representing a wave group. Since the Fourier components only correspond to the wave group, the wave group repeats continuously in time and space.

\begin{figure}[H]
	\centering
	\includegraphics[width=0.7\textwidth]{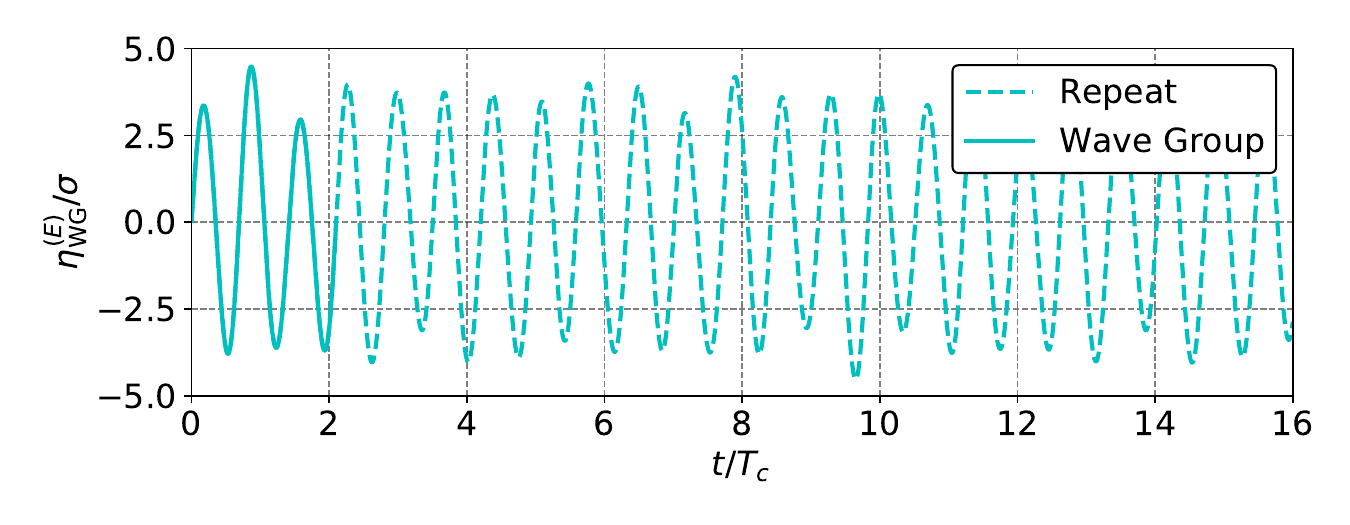}
	\caption{Encountered wave elevation traveling through the repeating wave group wave field with constant speed and heading.}
	\label{fig:ic_etaWGE}
\end{figure}

An estimate of encountering the wave group in a constant moving frame can be made by utilizing the deep water dispersion relation for the wavenumber, $k_{g_f} = \omega_{g_f}^2/g$, along with $U_e$ to transform the wave group time-history at a single point into a wave train that repeats in space and time. $g$ here refers to the acceleration due to gravity. This modifies Equation (15) to describe the wave group in the estimated encounter frame:

\begin{equation}
\eta_{\rm WG}^{\rm (E)} \left( t \right) \  = \ \sum_{f}^{} a_{g_f} cos \left( \omega_{g_f} t \ - \ k_{g_f} ( \cos(\mu) U_et) \ + \ \phi_{g_f} \right)
\label{eq:ic_wgE}
\end{equation}

where $\mu$ is the wave heading defined such that 180~deg is head seas, 0~deg is following seas, and 90~deg is starboard beam seas. Fig.~\ref{fig:ic_etaWGE} shows the encountered wave field as a vessel travels through the origin at $t=0$ with constant speed and heading. 

The wave group of interest is encountered over a different time-interval due to the forward speed of the vessel. Like the wave group elevation at the origin, the encountered wave group is repeated in time with the Fourier representation.

To create a physically realizable extreme event, the deterministic wave group in the encounter frame $\eta_{\rm WG}^{\rm(E)}(t)$ is embedded into the encountered irregular prelude $\eta_{\rm IP}^{\rm(E)}(t)$. The encountered wave group $\eta_{\rm WG}^{\rm(E)}(t)$ is shifted by $t_e$ such that the group’s initial upcrossing starts at location $x_e$, thus ensuring that the encountered wave group  $\eta_{\rm WG}^{\rm(E)}(t)$ will start directly at $t_e$. The new composite seaway in the estimated encounter frame $U_e t$ is formed with the blending functions $\beta_1$ and $\beta_2$ as:

\begin{equation}
\eta_{\rm C}^{\rm(E)}(t) = \left(1 - \beta_2\right)\left[(1-\beta_1)\eta_{\rm IP}^{\rm(E)}(t) + \beta_1\eta_{\rm WG}^{\rm(E)}(t - t_e)\right]
+ \beta_2\eta_{\rm IP}^{\rm(E)}(t)
\label{eq:ic_comp}
\end{equation}

where each blending function is defined as:

\begin{equation}
	\beta \ = \frac{1}{2}\left(1 + \tanh\left( \frac{t \ - \ t_e}{t_o} \right) \right)
	\label{eq:ic_blend}
\end{equation}

\begin{figure}[H]
	\centering
	\includegraphics[width=0.9\textwidth]{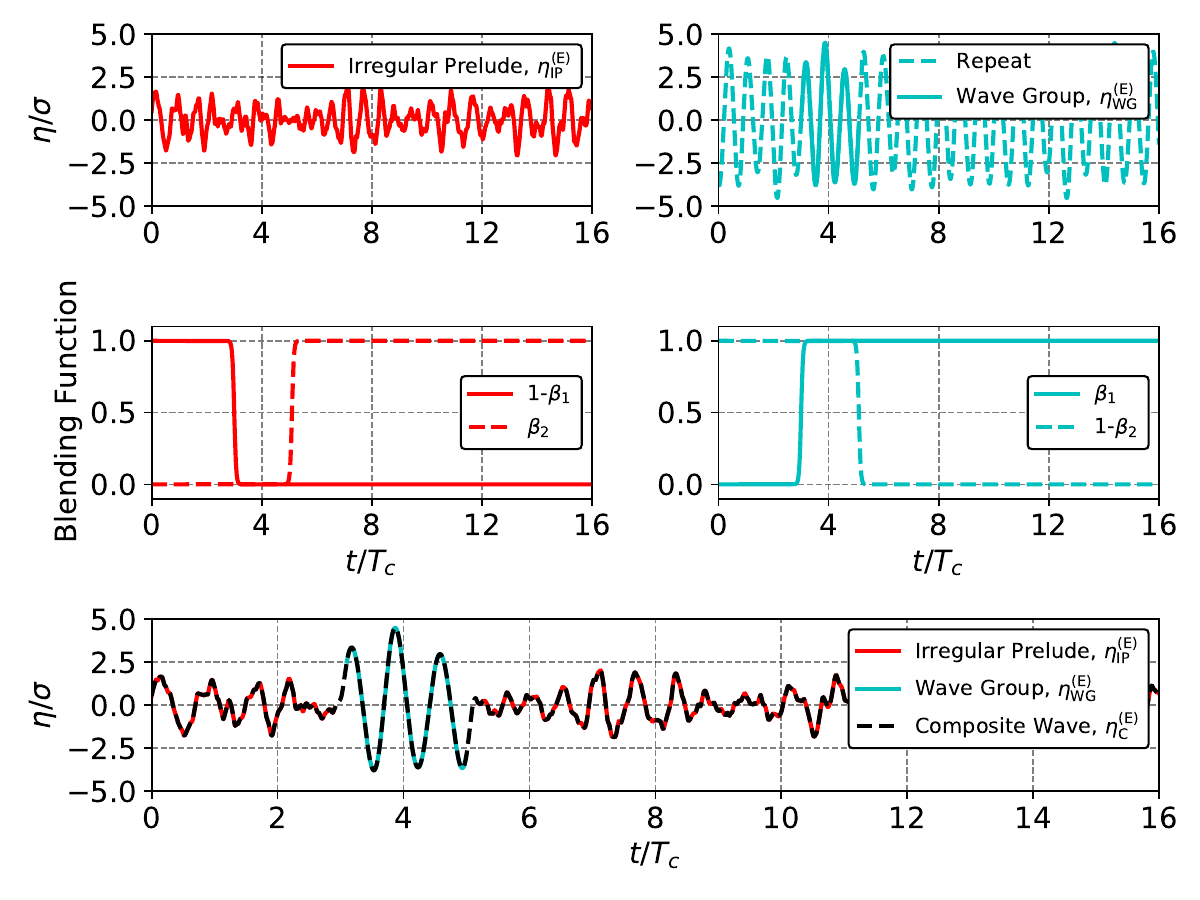}
	\caption{Formation of a composite wave by embedding a deterministic wave group into the irregular prelude in the estimated encounter frame.}
	\label{fig:ic_blend}
\end{figure}

The functions $\beta_1$ and $\beta_2$ correspond to the blending at the start and end of the wave group, respectively. Fig.~\ref{fig:ic_blend} shows the blending process for embedding the wave group into an irregular wave train to create a single composite wave train. The wave elevation and time in Fig.~\ref{fig:ic_blend} are non-dimensionalized by the standard deviation of the height from specified wave spectrum $\sigma$ and the period of the largest wave in the group $T_c$, respectively.

The time scale $t_o$ is selected with Eqn.~\eqref{eq:to} where the factor of 0.9 corresponds to approximately 95\% of the first signal at the start of the interval and 95\% of the second signal at the end. Eqn.~\eqref{eq:to} results in a composite wave, where the majority of the blending process occurs within two time intervals of duration $T_p/5$. To form the full composite wave train, $t_o$ is the same for $\beta_1$ and $\beta_2$, while $t_e$ depends on the start and end of the wave group. The portion of the composite wave train after the wave group is not considered when assessing the extreme ship response, but is required for the wave generation to ensure that the wave group sequences of interest are not repeated in the observed time.

\begin{equation}
	t_o \ = \ \frac{T_p}{10 \cdot \rm tanh^{-1}(0.9)}
	\label{eq:to}
\end{equation}

\begin{figure}[H]
	\centering
	\includegraphics[width=0.6\textwidth]{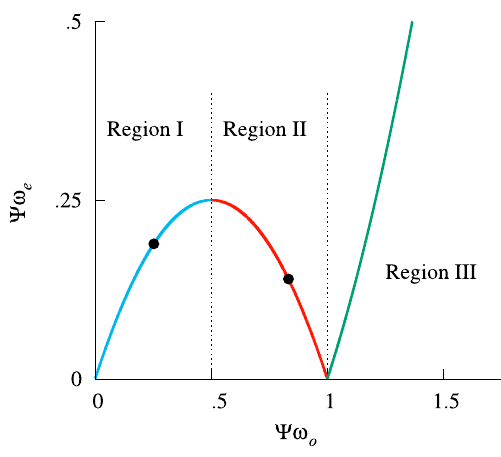}
	\caption{Relationship between encounter ($\omega_e$) and absolute ($\omega_o$) frequencies \citep{Xu2020}.}
	\label{fig:ic_encounterToAbs}
\end{figure}

The blending procedure outlined in Fig.~\ref{fig:ic_blend} produces a description of the composite wave train in the estimated encounter frame $\eta_{\rm C}^{\rm(E)}(t)$. To generate the necessary waves for a simulation or experiment, a full spatial and temporal description of the wave field is required. Therefore, $\eta_{\rm C}^{\rm(E)}(t)$ must be transformed to the earth-fixed frame. In beam to head seas (90-180~/180-270~deg), the transformation is straight forward utilizing the relationship $\omega_e=\omega_o-\Psi\omega_o^2$, where $\Psi=\cos(\mu)U_e/g$. With the conversion of the frequencies to the absolute frame and calculation of the wavenumber through the dispersion relation, the resulting composite wave train $\eta_{\rm C}^{\rm(E)}(\mathbf{x},t)$ is a function of both space and time. However, in beam to following seas (0-90,~270-360~deg), the Doppler effect causes the transformation to be multi-valued due to the movement of the ship relative to the direction of the waves. Fig.~\ref{fig:ic_encounterToAbs} illustrates the 3-to-1 mapping problem, where under the right conditions, an encounter frequency $\omega_e$ can correspond to three separate absolute frequencies $\omega_o$. \cite{Nielsen2017} introduced an algorithm to address this issue. In cases where this multi-valued problem exists, the encounter frequency can be mapped to the three separate absolute frequencies. A scale factor is then applied to the original corresponding amplitude Fourier components, based on a nominal wave spectrum estimated from the Fourier components in the encounter frame.

\begin{figure}[H]
	\centering
	\includegraphics[width=0.7\textwidth]{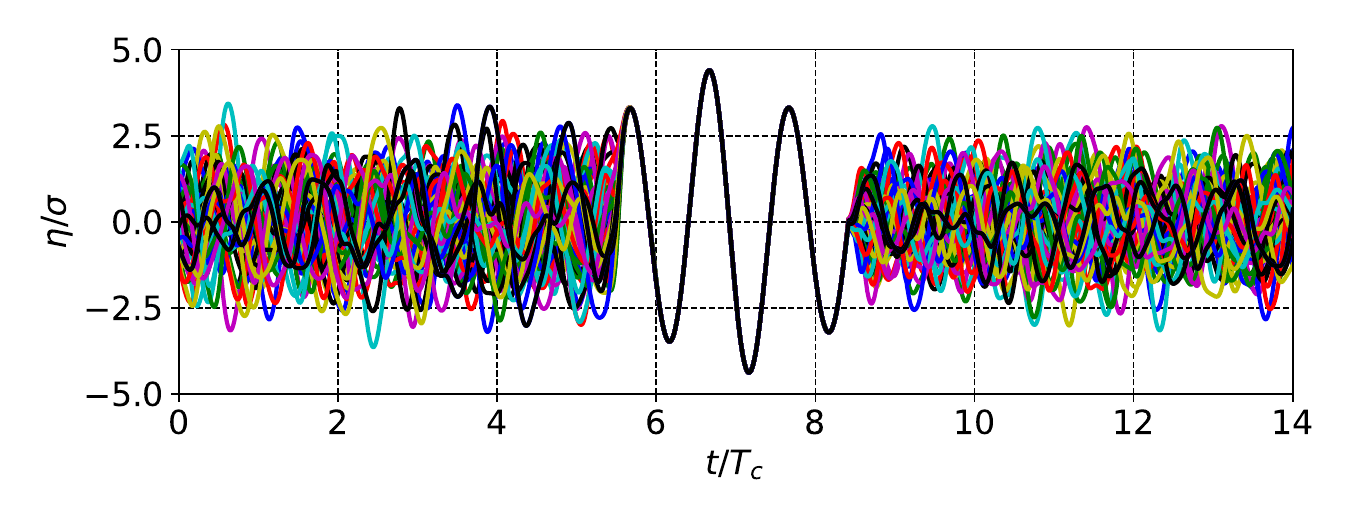}
	\caption{Ensemble of various waves with corresponding different irregular preludes for the same wave group that is shifted in time such that the largest wave in the group's peak occurs at the same time.}
	\label{fig:ic_ensemble_Hc16.875_Tc15}
\end{figure}

The natural initial condition method allows for enforcement of the encounter conditions of interest and generation of the wave group in a natural manner without intrusive and nonphysical measures or mechanisms (physical or virtual). All of the necessary information needed for the observation is contained within the composite wave train. Fig.~\ref{fig:ic_ensemble_Hc16.875_Tc15} reflects an ensemble of composite waves that possess the same wave group with different irregular preludes. The time-histories in Fig.~\ref{fig:ic_ensemble_Hc16.875_Tc15} have been shifted such that the peak of the largest wave in the group occurs at the same time for the composite wave trains to illustrate the methodology. All of the irregular preludes must account for the ramping up of the wave generation and the vessel reaching it's target speed. This logic ensures that vessel will reproduce the previously observed results that led to the encounter condition of interest.

An added benefit of the natural initial condition method is that a separate set of irregular wave trains that satisfy the same encounter condition can be determined. This identification allows for the construction of an ensemble of composite wave trains that can be studied to further understand sensitivity to encounter conditions. For instance, if only roll and roll velocity are considered for the encounter condition, but sway velocity contributes significantly, then irregular preludes with different sway velocities can be found to assess the importance of sway. Conversely, if unfavorable encounter conditions are selected such as heave motion and surge velocity when considering large roll, recognition of unsuitable quantities would be simple. Significant differences would occur in the ship response for the same encounter condition and wave group. This aspect of the natural initial condition methodology yields greater utility with more complicated failure mechanisms like capsizing due to broaching-to, where the quantities considered for the encounter conditions are not evident.

\section*{Case Study}

The presented methodology for modeling extreme ship motions with the CWG method for free-running vessels is demonstrated with simulations performed in LAMP for the David Taylor Model Basin (DTMB) 5415 hull form. The current work utilizes the LAMP-3 formulation, where the hydrodynamics (radiation and diffraction) is solved about the mean wetted surface (body-linear), and the hydrostatics and Froude-Krylov forces are solved over the instantaneous wetted surface (body-nonlinear). The blended nonlinear methodology can resolve a significant portion of nonlinear effects in most ship-wave problems at a fraction of the computational effort for the general body-nonlinear formulation and allows for large lateral motions and simulations of free-running vessels. 

Table~\ref{tab:5415properties} lists the loading condition and fluid properties for the DTMB 5415 case study. The loading condition is derived from CFD validation studies performed for the 5415M in \cite{Sadat-Hosseini2015}, while the fluid properties represent seawater at 20° \citep{ITTC2011}. The DTMB 5415 is free to surge, sway, heave, roll, pitch and yaw in the LAMP simulations. The vessel’s forward speed is controlled with a quasi-steady propeller performance model from \cite{Lee2003} and the rudders are modelled as low-aspect ratio foils that are actuated by a proportional-integral-derivative (PID) controller to maintain heading.

Table~\ref{tab:runmatrix} summarizes the seaway and operating conditions considered in the current work. A database of wave groups is constructed for Sea State 7 long-crested seas described by the JONSWAP spectrum \citep{Hasselmann1973}. A speed of 20~knots and heading of 90~deg (beam seas) is considered in this study. For the CWG method, roll was selected as the quantity of interest with the roll and roll velocity quantities as the encounter conditions. Other quantities of interest such as sway velocity could also have been included as encounter conditions. However, previous research with the CWG method has shown that only considering roll and roll velocity is sufficient to predict accurate probabilities of exceedance.

\begin{figure}[H]
	\centering
	\includegraphics[width=0.5\textwidth]{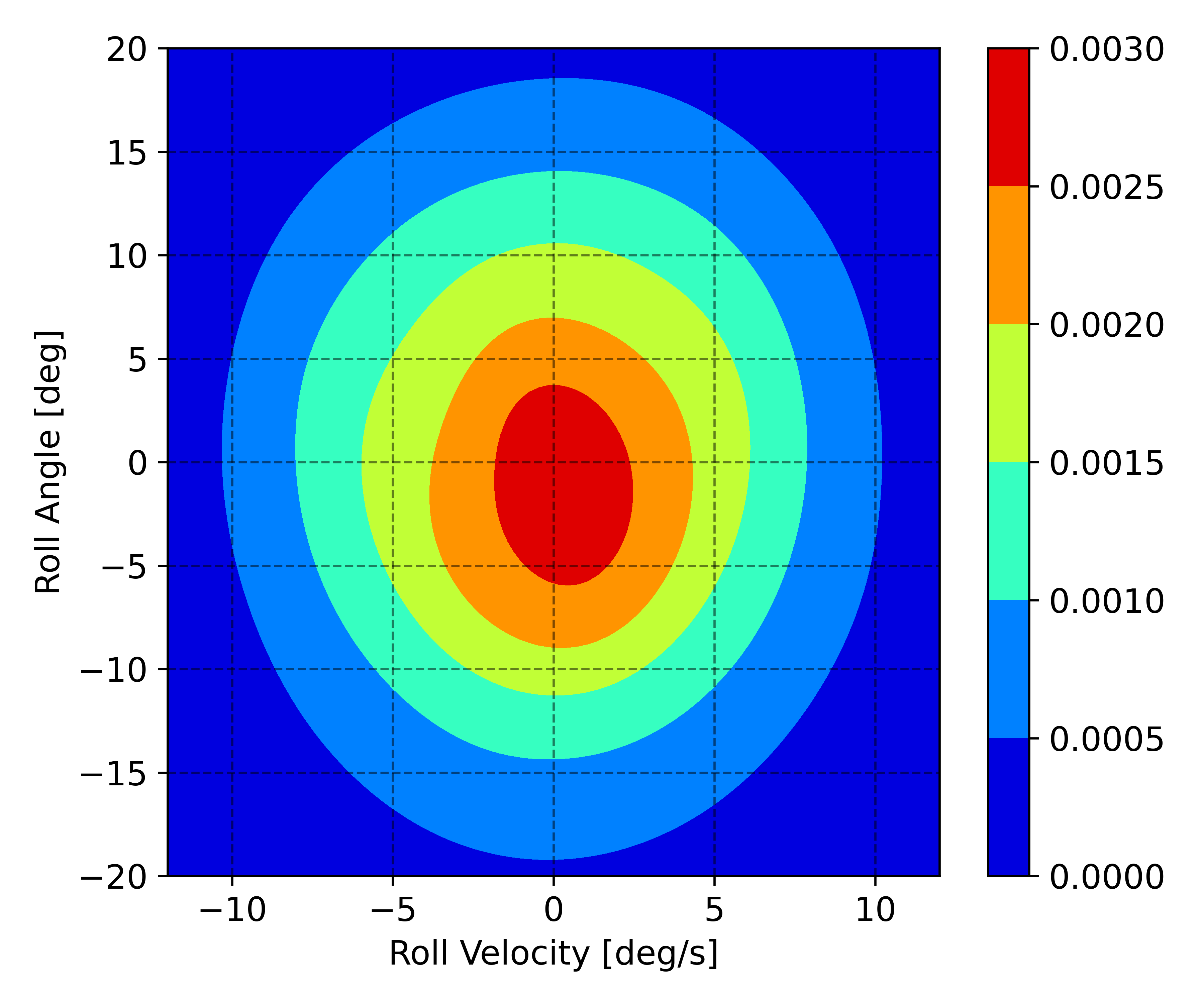}
	\caption{Probability distribution of encounter conditions.}
	\label{fig:probIC}
\end{figure}

\begin{figure}[H]
	\centering
	\includegraphics[width=0.5\textwidth]{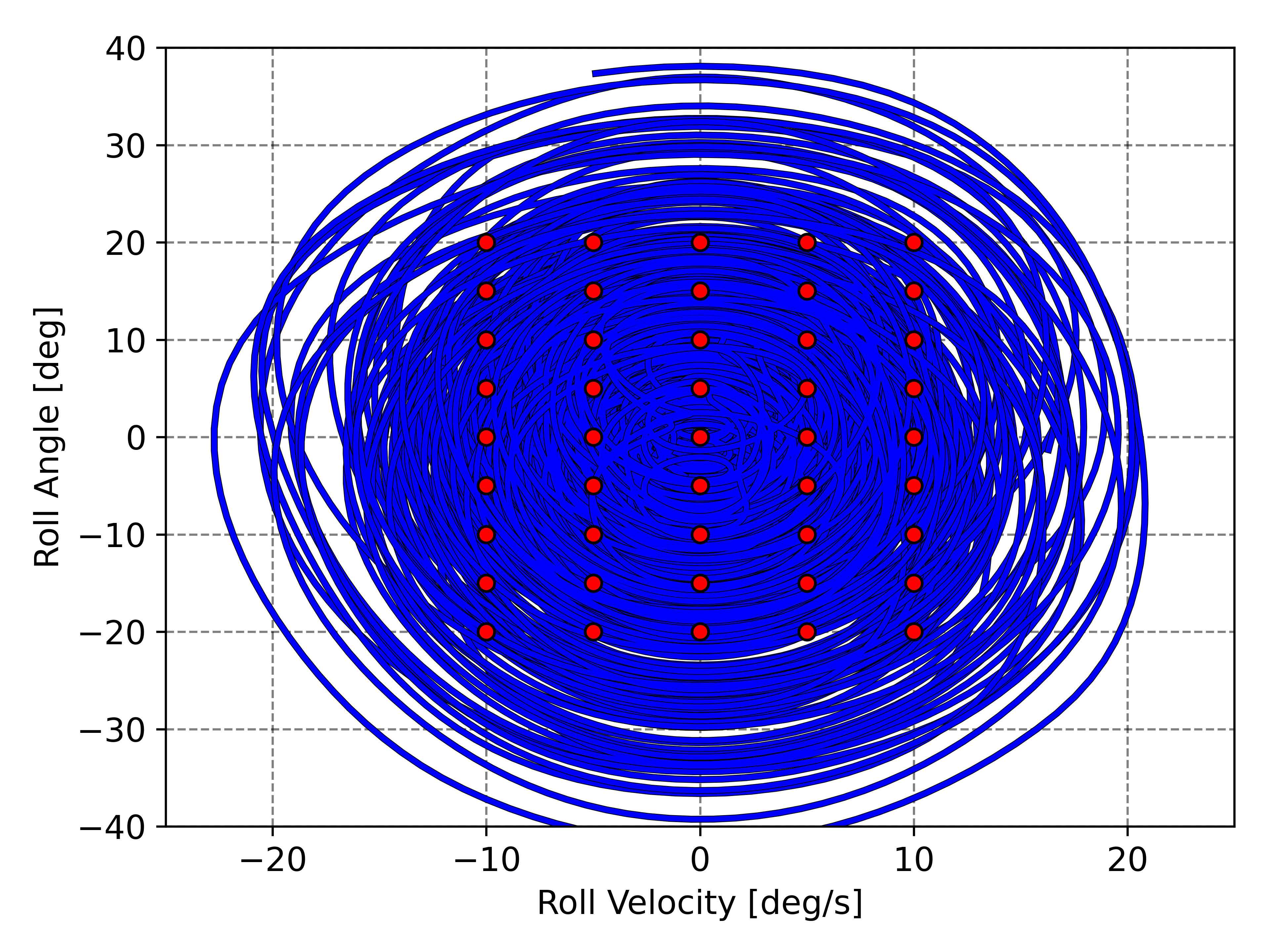}
	\caption{Selected encounter conditions from random irregular waves.}
	\label{fig:stateSpaceIC}
\end{figure}

Two hours of random irregular wave simulations were performed to identify the wave trains utilized for the natural initial condition as well as the probability of the encounter conditions. Fig.~\ref{fig:probIC} shows the joint probability distribution of roll angle and velocity from the two hours of random irregular waves. Fig.~\ref{fig:stateSpaceIC} displays the time traces of roll angle and velocity from the two hours random irregular wave simulations (20~realizations) and the selected encounter conditions. The 45~encounter conditions were uniformly spaced in both roll angle and roll velocity, respectively, so that the encounter condition pairs covered the majority of the joint probability distribution. Irregular preludes were selected from the time traces and the wave groups were embedded at the moment the encounter condition of interest occurred.

For this case study, wave groups were constructed for $j \leq 3$, $T_c$ ranging from 13 to 17~s in 1~s increments, and $H_c$ ranging from 8 to 21~m in 1~m increments. All of the wave groups and encounter conditions were combined to create a complete set of composite wave trains. LAMP simulations were performed for all the composite wave trains. For each threshold, the roll angle was interpolated with respect to $H_c$ to identify a value of $H_c$ that resulted in an exceedance of the various thresholds of interest.

\begin{table}[htbp!]
    \caption{Loading condition and fluid properties for the DTMB 5415 LAMP simulations.}
    \begin{center}
        \label{tab:5415properties}
        \begin{tabular}{l | c | l}
            \hline
            Properties & Units & Value \\
            \hline
            Length Between Perpendiculars   &    m  &   142.0\\
            Beam   &   m   &   19.06\\
            Draft   &   m   &   6.15\\
            Displacement   &   tonnes   &   8431.8\\
            Longitudinal Center of Gravity (+Fwd of AP)   &   m   &   70.317\\
            Vertical Center of Gravity (Above BL)   &   m   &   7.51\\
            Transverse Metacentric Height   &   m   &   1.95\\
            Roll Gyradius   &   m   &   7.62\\
            Pitch Gyradius   &   m   &   35.50\\
            Yaw Gyradius   &   m   &   35.50\\
            Density of Water   &   kg/m$^3$   &   1024.81\\
            Kinematic Viscosity of Water   &   m$^2$/s   &   1.0508e-6\\
            Accel. due to Gravity   &   m/s$^2$   &   9.80665 \\
            \hline	
        \end{tabular}
    \end{center}
\end{table}

\begin{table}[htbp!]
    \caption{Operating and seaway conditions for the DTMB 5415 case study.}
    \begin{center}
        \label{tab:runmatrix}
        \begin{tabular}{l | c | l}
            \hline
            Properties & Units & Value \\
            \hline
            Speeds   &   knots   &   20\\
            Headings   &   deg   &   90\\
            Sea State   &   -   &   7\\
            Wave Spectrum   &      &   JONSWAP\\
            Significant Wave Height, $H_s$   &   m   &   9.0\\
            Peak Modal Period, $T_p$   &   s   &   15\\
            Individual Run Length   &   s   &   360\\
            Validation Dataset   &   hr   &   1,000\\
            \hline	
        \end{tabular}
    \end{center}
\end{table}

\begin{figure}[H]
	\centering
	\includegraphics[width=0.5\textwidth]{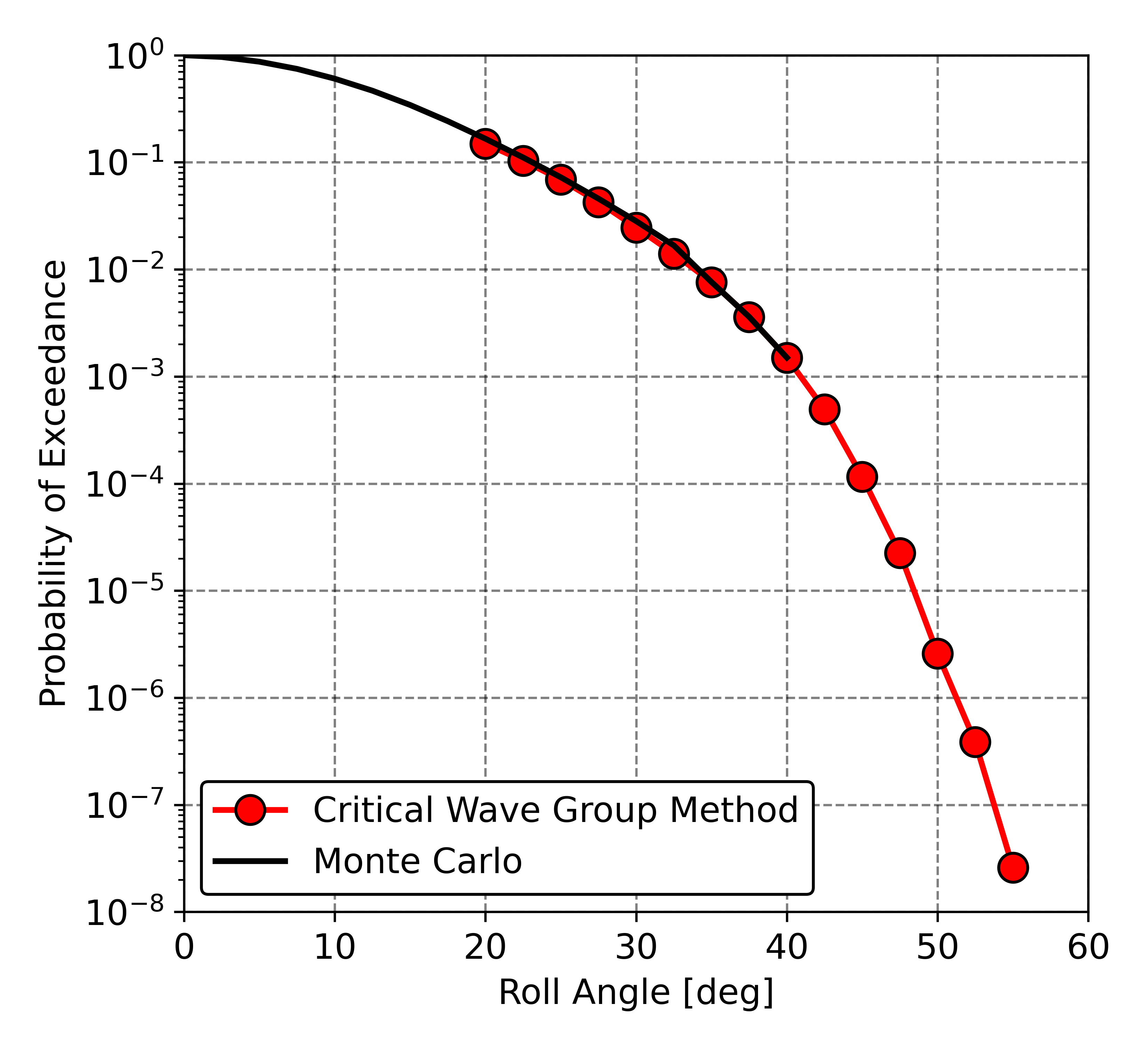}
	\caption{Probability of exceedance of roll in Sea State 7.}
	\label{fig:probExceed}
\end{figure}

Fig.~\ref{fig:probExceed} shows comparison of the probability of exceedance at various roll angle thresholds for the free running implementation of the CWG method and 1,000~hours of random irregular wave simulations. The probability of exceedance predicted by the free-running CWG method implementation are represented as red circles connected with a solid red line, and the 1,000~hours of random irregular wave simulations are denoted with a solid black line. Overall, the probabilities calculated with the CWG method represent the Monte Carlo dataset accurately for the thresholds exceeded within the 1,000~hours.

This is the first step towards validating the free-running implementation of the CWG method and still requires further research. A much larger validation dataset is needed as well as an extension to other speeds and headings before declaring that the CWG implementation can satisfactorily reproduce the probability of exceedance of roll for a free-running vessel.

\section*{Discussion}
The results presented in the case study demonstrate a strong agreement between the probability of exceedance predicted by the free-running CWG method and the benchmark 1,000-hour Monte Carlo simulation. This alignment is significant, as it suggests that the \emph{natural initial condition} methodology effectively captures the salient physics governing extreme roll events in beam seas. The probabilistic framework appears to be correctly weighting the contributions of both the critical wave group's structure and the vessel's dynamic state at the moment of encounter. The success of this limited validation implies that the method is not just a theoretical construct but a practical tool capable of identifying and leveraging the most impactful ship-wave interactions to predict rare events efficiently.

However, it is important to acknowledge the limitations inherent in the LAMP-3 formulation used for the simulations. Although this blended approach resolves the critical body-nonlinear Froude-Krylov and hydrostatic forces, which are dominant in driving large-amplitude roll, it does not capture fully nonlinear effects in wave radiation and diffraction, nor does it model viscous effects. Phenomena such as significant vortex shedding from bilge keels or the influence of wave breaking on the hull are therefore approximated. For the case of extreme roll in beam seas, the agreement shown in Fig.~\ref{fig:probExceed} suggests that the most critical nonlinearities are well represented. Nevertheless, the extension of this validation to scenarios where other nonlinearities become dominant remains an essential future step.

From a practical standpoint, these findings represent a significant move towards a computationally tractable framework for assessing extreme event probabilities in the early stages of vessel design or for operational guidance. Naval architects could employ this method to evaluate the performance of different hull forms or to establish safety envelopes without resorting to brute-force Monte Carlo simulations that are often prohibitively time-consuming and computationally expensive. The ability to quantify the probability and observe the physical event provides a dual benefit that is crucial to understanding and mitigating risk.

\section*{Conclusion}

This paper details a free-running implementation of the CWG method for the prediction of extreme roll. A description of the probability calculations and wave group construction within the CWG method are provided, as well as the extension of the methodology for free-running vessels utilizing the natural initial condition concept. A limited validation of the free-running implementation was performed for a single speed and sea state in beam seas, which showed that the probability calculations from the CWG method accurately represent the predictions from 1,000 hours of random irregular waves.

This paper represents the first step in formally validating the free-running implementation of the CWG method as proposed in \cite{Silva2022snh} and \cite{Silva2023}. The next steps will be to extend the validation dataset to longer exposure windows in order to validate the probability estimates for more extreme, lower-probability events. It is also critical to incorporate different speeds and headings—especially stern-quartering seas, which are known to induce complex and dangerous dynamic roll behavior. Success in these conditions would prove the framework's robustness for some of the most challenging operational scenarios.

To enhance the practicality of the framework, future work will also investigate the use of long short-term memory (LSTM) neural networks as surrogate models. This could drastically reduce the computational cost by replacing the need for a full time-domain simulation for every wave group and encounter condition pair, making the method accessible for rapid, iterative design evaluations. Finally, a comprehensive framework must include the uncertainty in the deterministic wave groups and the encounter conditions, as proposed in \cite{Silva2023}. Quantifying this uncertainty will provide essential confidence bounds on the final probability estimates, making the framework more robust for practical design and operational decisions. Accomplishing these steps will produce a methodology for estimating extreme events that is not only complete but is also a practical tool that addresses a critical gap in the ability to both quantify and observe extreme events for free-running vessels realistically.

\section*{Acknowledgments}
This work is supported by the Department of Defense (DoD) Science, Mathematics, and Research for Transformation (SMART) scholarship, the Naval Surface Warfare Center Carderock Division (NSWCCD) Extended Term Training (ETT), and the NSWCCD Naval Innovative Science and Engineering (NISE) programs. The authors would also like to acknowledge and thank the Office of Naval Research for the support of this work under contracts N00014-20-1-2096 by the program manager Woei-Min Lin. 
	
	\bibliographystyle{abbrvnat}
	\bibliography{References}  
	
\end{document}